# WEDGE ABSORBERS IN FINAL COOLING FOR A HIGH-ENERGY HIGH-LUMINOSITY LEPTON COLLIDER*

David Neuffer, Fermilab, Batavia, IL 60510 USA,
D. Summers, University of Mississippi, Oxford, MS 38655 USA
P. Snopok and T. Mohayai, IIT, Chicago, IL 60166 USA

*Abstract*

A high-energy muon collider scenario requires a final cooling system that reduces transverse emittance to ~25 microns (normalized) while allowing longitudinal emittance increase. Ionization cooling using high-field solenoids (or Li Lens) can reduce transverse emittances to ~100 microns in readily achievable configurations, confirmed by simulation. Passing these muon beams at ~100 MeV/c through cm-sized diamond wedges can reduce transverse emittances to ~25 microns, while increasing longitudinal emittances by a factor of ~25. Implementation will require optical matching of the exiting beam into downstream acceleration systems.

## INTRODUCTION

In muon collider scenarios, emittances as small as 25μ (transverse, rms, normalized) are required to ensure high luminosity at multiTeV energies [1]. Ionization cooling is used to reduce transverse emittances, following the cooling equation:

$$\frac{d\varepsilon_N}{ds} = -\frac{g_t}{\beta^2 E}\frac{dE}{ds}\varepsilon_N + \frac{\beta_\perp E_s^2}{2\beta^3 m_\mu c^2 L_R E} \quad (1)$$

where the first term is the frictional cooling effect and the second is the multiple scattering heating term. Here $L_R$ is the material radiation length, $\beta_\perp$ is the betatron focusing function, and $E_s$ is the characteristic scattering energy (~14 MeV), and $g_t$ is the transverse partition parameter. The equilibrium emittance is:

$$\varepsilon_{N,eq} = \frac{\beta_\perp E_s^2}{2\beta g_t m_\mu c^2 L_R \frac{dE}{ds}}.$$ At optimal cooling energies (~200 MeV), cooling at $\beta_t = $ ~1cm to $\varepsilon_{N,eq}= $ ~$10^{-4}$ m is relatively practical [2]. Cooling to smaller emittances requires cooling at low momentum with very high focusing fields, and, at low momentum, ionization cooling greatly increases energy spreads. The resulting systems reduce transverse emittances at the cost of increased longitudinal emittance, with the result that 6-D phase space emittance remains nearly constant[3].

Since this "final cooling" is predominantly emittance exchange, we propose that this can be done more efficiently by explicit emittance exchange techniques. Energy loss in a wedge absorber is a particularly promising one [4,5].

## WEDGE EXCHANGE FORMALISM

Figure 1 shows a stylized view of the passage of a beam with dispersion $\eta_0$ through a wedge absorber. The wedge is approximated as an object that changes particle momentum offset $\delta = \Delta p/P_0$ as a function of $x$, and the wedge is shaped such that that change is linear in $x$. (The change in average momentum $P_0$ is ignored, in this approximation. Energy straggling and multiple scattering are also ignored.) The rms beam properties entering the wedge are given by the transverse emittance $\varepsilon_0$, betatron amplitude $\beta_0$, dispersion $\eta_0$ and relative momentum width $\delta_0$. (To simplify discussion the beam is focussed to a betatron and dispersion waist at the wedge: $\beta_0'$, $\eta_0' = 0$. This avoids the complication of changes in $\beta'$, $\eta'$ in the wedge.) The wedge is represented by its relative effect on the momentum offsets $\delta$ of particles within the bunch at position $x$: $\frac{\Delta p}{p} = \delta \to \delta - \frac{(dp/ds)\tan\theta}{P_0}x = \delta - \delta'x$

$dp/ds$ is the momentum loss rate in the material ($dp/ds = \beta^{-1}dE/ds$). $x\tan\theta$ is the wedge thickness at transverse position $x$ (relative to the central orbit at $x=0$), and $\delta' = dp/ds \tan\theta/P_0$ to indicate the change of $\delta$ with $x$.

Under these approximations, the initial dispersion and the wedge can be represented as linear transformations in the $x$-$\delta$ phase space projections and the transformations are phase-space preserving. The dispersion can be represented by the matrix: $\mathbf{M}_\eta = \begin{bmatrix} 1 & \eta_0 \\ 0 & 1 \end{bmatrix}$, since $x \Rightarrow x + \eta_0 \delta$. The wedge can be represented by the matrix: $\mathbf{M}_\delta = \begin{bmatrix} 1 & 0 \\ -\delta' & 1 \end{bmatrix}$, obtaining $\mathbf{M}_{\eta\delta} = \begin{bmatrix} 1 & \eta_0 \\ -\delta' & 1-\delta'\eta_0 \end{bmatrix}$. Writing the $x$-$\delta$ beam distribution as a phase-space ellipse: $g_0 x^2 + b_0 \delta^2 = \sigma_0 \delta_0$, and transforming the ellipse by standard betatron function transport techniques obtains new coefficients $b_1$, $g_1$, $a_1$, which define the new beam parameters[6]. The momentum width is changed to:

$$\delta_1 = \sqrt{g_1 \sigma_0 \delta_0} = \delta_0 \left[ (1-\eta_0 \delta')^2 + \frac{\delta'^2 \sigma_0^2}{\delta_0^2} \right]^{1/2}.$$

The bunch length is unchanged. The longitudinal emittance has therefore changed simply by the ratio of energy-widths, which means that the longitudinal emittance has changed by the factor $\delta_1/\delta_0$. The transverse emittance has changed by the inverse of this factor:

$$\varepsilon_1 = \varepsilon_0 \left[ (1-\eta_0 \delta')^2 + \frac{\delta'^2 \sigma_0^2}{\delta_0^2} \right]^{-1/2}$$ The new values of ($\eta$,

---

*Work supported by by FRA Associates, LLC under DOE Contract No. DE-AC02-07CH11359.
#neuffer@fnal.gov

$\beta$) are:

$$\eta_1 = -\frac{a_1}{g_1} = \frac{\eta_0(1-\eta_0\delta') - \delta'\frac{\sigma_0^2}{\delta_0^2}}{(1-\eta_0\delta')^2 + \delta'^2\frac{\sigma_0^2}{\delta_0^2}} \text{ and}$$

$$\beta_1 = \beta_0\left[(1-\eta_0\delta')^2 + \frac{\delta'^2\sigma_0^2}{\delta_0^2}\right]^{-1/2}.$$

Note that the change in betatron functions ($\beta_1$, $\eta_1$) implies that the following optics should be correspondingly matched. A single wedge exchanges emittance between one transverse dimension and longitudinal; the other transverse plane is unaffected. Serial wedges could be used to balance $x$ and $y$ exchanges, or a more complicated coupled geometry could be developed.

Wedge parameters can be arranged to obtain large exchange factors in a single wedge. In final cooling we wish to reduce transverse emittance at the cost of increased longitudinal emittance.

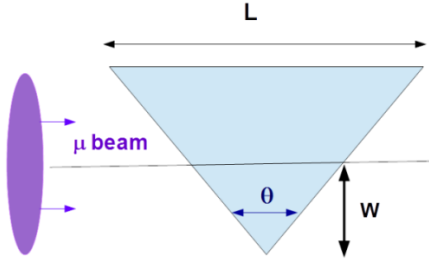

Figure 1: Schematic view of a muon beam passing through a wedge.

## WEDGES FOR FINAL COOLING

For final cooling, the beam and wedges should be matched to obtain a large factor of increase in momentum spread. That means that the energy spread induced by the wedge should be much greater than the initial momentum spread: $\delta_0 \ll \delta'\sigma_0 = \frac{2\tan\left(\frac{\theta}{2}\right)\frac{dp}{ds}}{P_0}\sigma_0$. Thus the incident beam should have a small momentum spread and small momentum $P_0$ and the wedge should have a large $\tan(\theta/2)$, large $dp/ds$ and a large $\sigma_0 = (\varepsilon_0\beta_0)^{1/2}$. ($\varepsilon_0$ is unnormalized, rms in this section.) Beam from a final cooling segment (high-field solenoid or Li lens) is likely to have $P_0 \approx 100$—150 MeV/c, and $\delta p \approx 3$ MeV/c. For optimum single wedge usage, $\delta p$ should be reduced to ~0.5MeV/c, and this can be done by rf debunching of the beam to a longer bunch length. To simplify initial exploration, the dispersion entering the wedge is set to zero ($\eta_0 = 0$), although the exchange can be improved by matching to ($\eta_0 = 1/\delta'$). The best material is a high-density low-Z material (Be or C (diamond density) or $B_4C$ (almost as good)). Emittance change passing through a wedge was simulated using ICOOL, and results in good agreement with the above transport model are obtained.

Table 1: Beam parameters at entrance, center and exit of a w=3mm, $\theta$=85° diamond wedge. ($z$ = 0, 0.6, 1.2cm) The 0.6cm values can be obtained with a half-size wedge.

| z (cm) | $P_z$(MeV/c) | $\sigma_E$(MeV) | $\varepsilon_x$ ($\mu$) | $\varepsilon_y$($\mu$) | $\varepsilon_z$ (mm) |
|---|---|---|---|---|---|
| 0 | 100 | 0.5 | 129 | 127 | 1.0 |
| 0.6 | 95.2 | 2.0 | 40.4 | 130 | 4.0 |
| 1.2 | 90.0 | 3.9 | 25.0 | 127 | 7.9 |

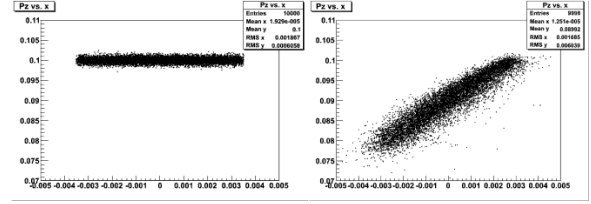

Figure 2: x-P projections of beam before and after the wedge.

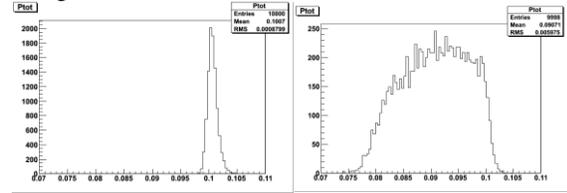

Figure 3: Momentum spread distributions before and after a final cooling wedge.

## FINAL COOLING WITH WEDGES

A final cooling scenario using as few as 2 wedges can be developed. The sequence could be:

1. Transverse Cooling. A cooling system to minimize emittances within reasonable fields is used. It should cool $\varepsilon_x$ and $\varepsilon_y$ to ~1.3×10$^{-4}$m, while $\varepsilon_L\approx$~0.003m. This could be the initial sector of the baseline front end.

2. Match into the first wedge: The beam is stretched to $\sigma_z$ = ~0.6m to enable phase-energy rotation to $\delta E <$ 0.5 MeV while being decelerated to ~100 MeV/c. A focus onto the first wedge causes an emittance exchange to $\varepsilon_x$ = 25$\mu$, $\varepsilon_y$ = 130$\mu$, $\varepsilon_L$ =~0.015m .

3. Match into second wedge: The beam is stretched to $\sigma_z$= ~3m to enable phase space rotation to $\delta E <$ 0.5 MeV while reaccelerating to ~100 MeV/c. A second wedge obtains $\varepsilon_x$ = 30$\mu$, $\varepsilon_y$ = 25$\mu$, $\varepsilon_L$ =~0.075m.

4. The beam is phase-energy rotated and accelerated and bunched as a 12m long bunch train (12 bunches at 300 MHz or 24 at 600 MHz).

5. Longitudinal recombination. The bunches are accelerated into a ring that combines them by snap coalescence into a single bunch ($\varepsilon_x$ < 30$\mu$, $\varepsilon_y$ < 30$\mu$, $\varepsilon_L$ =~ 0.075m).

## HEATING EFFECTS

In the initial matrix approximation, both the cooling and heating effects in eq. 1 are ignored. To first approximation this effect is estimated by:

$$\Delta\varepsilon_N\big|_{material} = -\frac{1}{\beta^2 E}\frac{dE}{ds}\varepsilon_N \Delta z + \frac{\beta_\perp E_s^2}{2\beta^3 m_\mu c^2 L_R E}\Delta z,$$

where $\Delta z$ is the width of the center of the wedge. To minimize heating $\beta_\perp$ should be relatively small (< ~3 cm).

The divergence in energy spread caused by the increase in energy loss with decreasing momentum is an important heating effect, which is larger at smaller momentum. It can be reduced by shaping the wedge to reduce the energy loss of lower energy particles.

## EXPERIMENT AT MICE PARAMETERS

The MICE experiment has considered inserting a wedge absorber into the beam line for measurements of emittance exchange cooling [7]. The layout would be a scale model of the final cooling wedge examples (~10× larger).

As an example we consider using a polyethylene ($C_2H_4$) absorber with w=5cm, θ=60°, with the wedge oriented along x. (A Be or LiH wedge would have superior performance, but greater expense, and would not greatly improve the initial proof of principle demonstration.). The incident beam would be matched to $\sigma_x$ = 2.5 cm, ($\varepsilon_t$= 3mm, $\beta_t$=36cm) $P_0$=200 MeV/c, corresponding to a baseline MICE beam setting [8], but with $\delta p$ = 2 MeV/c. The small $\delta p$ is obtained by software selection of the incident beam. This example obtains an increase in $\delta p$ by a factor of ~4 accompanied by a reduction in $\varepsilon_x$ by a factor of ~4.

This example was simulated in ICOOL [9], with results presented in table 2 and displayed in Fig. 4 and 5. The resulting scenario would be an interesting scaled model of a final cooling scenario and would test the basic physics and optics of the exchange configuration.

Table 2: Beam parameters at entrance, center and exit of a w=3 cm, θ=60° polyethelene wedge. (z = 0, 6, 12 cm).

| z (cm) | $P_z$ (MeV/c) | $\sigma_E$ (MeV) | $\varepsilon_x$ (mm) | $\varepsilon_y$ (mm) | $\varepsilon_z$ (mm) |
|---|---|---|---|---|---|
| 0 | 200 | 1.8 | 3.0 | 3.0 | 2.9 |
| 6 | 193 | 3.9 | 1.44 | 3.0 | 6.8 |
| 12 | 182 | 8.6 | 0.76 | 3.0 | 14.3 |

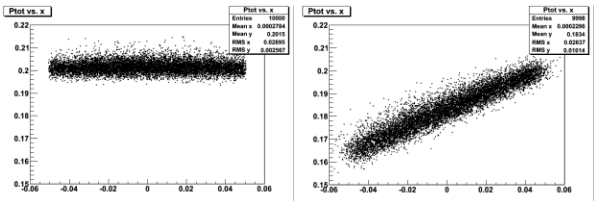

Figure 4: x-P projections of beam before and after the MICE wedge.

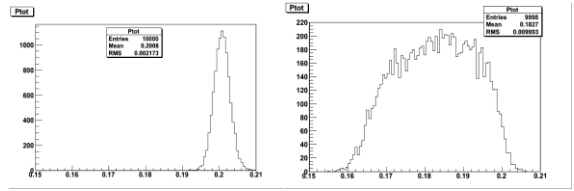

Figure 5: Momentum spread distributions before and after the MICE wedge. (Compare with fig. 3.)

Figure 5 displays the momentum spread distribution before and after the wedge. The MICE experiment can measure these accurately and that measurement would be a strong confirmation of the exchange effect. Note that this is a large effect, much larger than other cooling effects in the MICE beam. A more complete evaluation would evaluate 6-D emittance eigenvalues, properly corrected for dispersion. This is in principle possible but may be difficult within the MICE optics since the wedge introduces an x-y asymmetry and the optics into the spectrometer solenoid includes only solenoids, with radial focusing. Filamentation in the mismatched optics may obscure the result.

The same absorber, but with input beam selected to have dispersion and large $\delta p$, can also be used to demonstrate $\delta p$ reduction, as is needed for longitudinal cooling.

## CONCLUSION

Wedges at final cooling parameters can reduce the transverse emittance of muon beams to small values compatible with a high-luminosity high-energy lepton colliders. A scaled experiment demonstrating the principle can be performed at MICE.

## REFERENCES

[1] R. Palmer, "Muon Colliders", Rev. Acc. Sci. and Tech., **7**, 137 (2014).
[2] D. Neuffer, Part. Acc. **14**, 75 (1983).
[3] H. Sayed, R. Palmer and D. Neuffer, Phys. Rev. ST Accel. Beams **18**, 091001 (2015).
[4] D. Neuffer, H. Sayed, D. Summers, and T. Hart, Proc. IPAC15, TUBD2, 1384 (2015).
[5] D. Neuffer et al., submitted to J. Inst. (2016).
[6] D. Neuffer, Fermilab note MuCOOL003 (1996).
[7] C. Rogers, P. Snopok, L. Coney, A. Jansson, "Wedge Absorber Design for the Muon Ionisation Cooling Experiment", MICE 290 (2010).
[8] M. Bogommilov et al., J. Inst. **7**(5), P05009 (2012).
[9] R. Fernow et al. "ICOOL", Proc. 1999 PAC, New York, p. 3020, http://pubweb.bnl.gov/people/fernow/.